\documentclass[12pt,preprint]{aastex}

\def\Msun{$M_{\odot}$}
\def\Mvir{$M_{\rm VIR}$}
\def\Mdust{$M_{\rm dust}$}

\def\kms{kms$^{-1}$}
\def\vlsr{$V_{\rm LSR}$}
\def\HII{H{\sc ii}~}

\def\amm{NH$_3$}

\def\h13cop{H$^{13}$CO$^+$}
\def\hc5n{HC$_{5}$N}

\def\t32{$J=3-2$}
\def\h13cn{H$^{13}$CN}
\def\cc34s{CC$^{34}$S}
\def\n2hp{N$_2$H$^+$}

\def\13co{$^{13}$CO}
\def\c18o{C$^{18}$O}
\def\ch3cn{CH$_{3}$CN}
\def\c34s{C$^{34}$S}
\def\3423{$3_4-2_3$}

\def\Tmb{$T_{\rm mb}$}

\def\deg{\hbox{$^{\circ}$}}
\def\arcmin{\hbox{$^{\prime}$}}
\def\arcsec{\hbox{$^{\prime\prime}$}}

\def\cc{cm$^{-3}$}
\def\degree{$^{\circ}$}
\def\kms{km s$^{-1}$}

\def\tdust{$T_{\rm dust}$}
\def\tex{$T_{\rm ex}$}
\def\tmb{$T_{\rm mb}$}
\def\Tk{$T_{\rm k}$}

\def\vsix{$V_{\rm CO 6-5}$}
\def\vlsr{$V_{\rm LSR}$}
\def\hpbw{$\theta_{\rm HPBW}$}
\def\mrot{$M_{\rm rot}$}
\def\Mrot{$M_{\rm rot}$}
\def\Mvir{$M_{\rm VIR}$}
\def\Mlte{$M_{\rm LTE}$}
\def\h2{$H_{2}$}

\def\dvfwhm{$\Delta v_{\rm FWHM}$}
\def\jybm{Jy beam$^{-1}$}

\slugcomment{Accepted for publication in The Astrophysical Journal.}

\shorttitle{Submillimeter Observations of The IRAS 20126+4104 Clump} 
\shortauthors{Shinnaga et al.}

\begin{document}


\title{Submillimeter Observations of The Isolated Massive Dense Clump IRAS 20126+4104}

\author{Hiroko Shinnaga\altaffilmark{1},  
Thomas G. Phillips\altaffilmark{1,2}, 
Ray S. Furuya\altaffilmark{3}, and 
Riccardo Cesaroni\altaffilmark{4} 
} 
\email{shinnaga@submm.caltech.edu}


\altaffiltext{1}{California Institute of Technology Submillimeter Observatory (CSO)
   111 Nowelo St. Hilo HI 96720}
\altaffiltext{2}{
Division of Physics, Mathematics, and Astronomy, California Institute of Technology,
1200 East California Pasadena CA 91125}
\altaffiltext{3}{Subaru Telescope, National Astronomical Observatory of Japan, 
   650 N. A'ohoku Pl. Hilo HI 96720}
\altaffiltext{4}{Osservatorio Astrofisico di Arcetri, INAF, Largo E. Fermi 5, 50125 Firenze, Italy}


\begin{abstract}
We used the CSO 10.4 meter telescope to image the 350 $\mu$m and 450$\mu$m continuum 
and CO $J=6-5$ line emission of the IRAS 20126+4104 clump.  
The continuum and line observations show that the clump is isolated over a 4 pc region
and has a radius of $\sim$ 0.5 pc.  
Our analysis shows that the clump has a radial density profile $\propto r ^{-1.2}$ for $r \la$ 0.1 pc and 
has $\propto r^{-2.3}$ for $r \gtrsim$ 0.1 pc
which suggests the inner region is infalling, 
while the infall wave has not yet reached the outer region.   
Assuming temperature gradient of r$^{-0.35}$, the power law indices become 
$\propto r ^{-0.9}$ for $r \la$ 0.1 pc and $\propto r^{-2.0}$ for $r \gtrsim$ 0.1 pc. 
Based on a map of the flux ratio of 350$\mu$m/450$\mu$m, we identify three distinct regions:
a bipolar feature that coincides with the large scale CO bipolar outflow;
a cocoon-like region that encases the bipolar feature and has a warm surface; and 
a cold layer outside of the cocoon region.  
The complex patterns of the flux ratio map indicates that the clump is no longer uniform 
in terms of temperature as well as dust properties.  
The CO emission near the systemic velocity traces the dense clump and  
the outer layer of the clump shows narrow line widths ($\la$ 3 \kms).  
The clump has a velocity gradient of $\sim$ 2 \kms pc$^{-1}$, which 
we interpret as due to rotation of the clump, as the equilibrium mass ($\sim$ 200 \Msun) is comparable 
to the LTE mass obtained from the CO line.  
Over a scale of $\sim$ 1 pc, the clump rotates in the opposite sense with respect to the 
$\lesssim$ 0.03 pc disk associated with the (proto)star.  
This is one of four objects in high-mass and low-mass star forming regions 
for which a discrepancy between the rotation sense of the envelope and the core
has been found,
suggesting that such a complex kinematics may not be unusual 
in star forming regions.  
\end{abstract}



\keywords{ISM: clouds --- ISM: jets and outflows 
--- ISM: kinematics and dynamics --- ISM: lines and bands --- 
stars: formation } 



\section{INTRODUCTION}
High mass star formation, especially in its earliest phase, is poorly understood because
of the nature of the objects in this phase.  
First of all, young high mass stars are deeply embedded in their dense natal cloud cores 
and form and evolve on a short time scale, much shorter than that of low mass star formation. 
Moreover, once a massive star has reached the zero age main sequence, 
its strong UV radiation pressure
heavily affects the surrounding material, thus making it difficult
to trace back the initial conditions of the molecular cloud.
Finally, high mass stars form in clusters at relatively large distances from us
(typically a few kpc): it is hence not easy to resolve individual 
young stellar objects, even for the closest massive star 
forming region, Orion  (e.g., Plambeck et al. 1982, Greenhill et al. 1998, Beuther et al. 2005), 

The early B type massive (proto)star, IRAS 20126+4104, is a well-studied object 
which provides us the opportunity to
study the early phase of massive star formation 
in a relatively simple configuration.
This consists of a luminous IRAS source embedded in a dense core and associated
with both a bipolar outflow and a circumstellar Keplerian disk
(imaged in NH$_3$, \ch3cn and C$^{34}$S; Zhang et al. 1998, Cesaroni et al. 1997, 1999, 2005).
The Caltech Millimeter Array (OVRO-MMA) was used to image
the large scale ($\sim$ 0.5 pc) north-south outflow in detail using the CO $J=1-0$ transitions \citep{she00}.  
It appears that this N--S direction is not preserved on a smaller scale (0.1~pc) where a jet
oriented SE--NW has been imaged in a variety of tracers, such as
shocked H$_{2}$ (Cesaroni et al. 1997, Shephered et al. 2000),  \amm (3,3)/(4,4) emission (Zhang et al. 1999),  
SiO and HCO$^{+}$ emission (Cesaroni et al. 1997, 1999), 
and scattered light in the near infrared (NIR) continuum (e.g., Ayala et al. 1998, Sridharan et al. 2005).
The position angle ($PA$) of the jet is $\sim$ 150\deg, significantly different from that of the large scale molecular outflow
($PA\simeq -25$\deg; Shepherd et al. 2000, Lebr{\'o}n et al. 2006).  
Such a discrepancy has been interpreted as
precession of the jet/outflow axis (Shephered et al. 2000, Cesaroni et al. 2005).  

While many studies have been done on this object,
only few of these have focused on the dense pc-scale clump hosting the (proto)star.  
It is thus of interest to
investigate the physical properties and evolution of such a clump and its interaction
with the outflow from the (proto)star.  

Here we present a study of the dense clump and bipolar outflow 
based on submillimeter continuum and spectroscopic observations 
taken with the 10.4 meter Leighton telescope at the Caltech Submillimeter Observatory (CSO)
\footnote{The Caltech Submillimeter Observatory is supported by the 
NSF grant AST-0229008.}.  
Note that 1.5 kpc is used as the distance of the object to estimate physical parameters in this paper.  
A discussion regarding the distance of this object is given in the Appendix.  

\section{OBSERVATIONS AND DATA REDUCTION} \label{obs}
Submillimeter continuum and molecular line observations were carried out 
towards the massive dense clump 
(Table 1).  
The dust continuum observations at 350 $\mu$m and 450 $\mu$m were made on 2005 May 10 and 11 using the Submillimeter High Angular Resolution Camera II
(SHARC II, Dowell et al. 2003; Voellmer et al. 2003) 
mounted on a Nasmyth focus. 
Both the 350$\mu$m and 450$\mu$m beams have simple Gaussian shapes with 
half-power beam widths (\hpbw) of 8.7\arcsec~ and 9.8\arcsec~ (see Table\ref{tbl-1}).  
The 384 pixels of the bolometer array cover a region of about $\sim$ 1\arcmin~ $\times$ 2.6\arcmin~ at 350 $\mu$m.  
Pointing and focus were checked regularly. 
The pointing drift due to temperature \citep{sh04b}  was subtracted in the data reduction process.  
The final pointing accuracy is within $\sim$ 2\arcsec. 
The Dish Surface Optimization System (DSOS) \citep{leo05} was used during the observations: this allows us to apply a gravitational
deformation correction to the main mirror 
physically and is thus a key component for high sensitivity short submillimeter wave observations.
Zenith opacities $\tau_{225}\sim$0.06 at the wavelength of 1300 $\mu$m were measured with the CSO tau meter
during the 350$\mu$m observations, while during the acquisition of the 450$\mu$m data we measured $\tau_{225}\sim$0.09.
The data were reduced with the software Comprehensive Reduction Utility (CRUSH, Kovacs 2006)  
version 1.41.  The flux was calibrated with Neptune for both wavelengths, with an estimated 
calibration uncertainty of 7\% at 350$\mu$m and 5\% at 450$\mu$m.  
The estimated flux of the planet was 94.6 Jy at 350$\mu$m on 2005 May 10 and 55.9 Jy at 450$\mu$m 
on 2005 May 11.  Note that the angular size of the planet 
was much smaller than the beam sizes at these wavelengths.  
An area of $\sim$ 9\arcmin~ centered at the source position was mapped with the instrument, in box-scan mode. 
The RMS noise of the final maps is about 200 mJy/beam for both bands. 
After calibration, data analysis was done using the program GRAPHIC of the GILDAS software.  

Spectroscopic observations in the CO $J=6-5$ line ($\nu$= 691.473076 GHz, Goldsmith et al. 1981)
were made on 2004 July 8 and 9 
using a 690 GHz heterodyne receiver.  
An area of 2.\arcmin7 $\times$ 2.\arcmin7 was sampled with steps of 5\arcsec~ and 20\arcsec. 
while the innermost region of 1\arcmin~20\arcsec $\times$ 1\arcmin~ was sampled with steps of 5\arcsec. 
$\tau_{225}$ 
was between 0.07 and 0.085 during the observations. 
The SIS receiver operation at 4K produced typical single sideband system temperatures
(measured with the 1.5 GHz bandwidth Acousto-optic spectrometer)
between $\sim$ 7000 K and 10900 K at 691 GHz, for elevations in the range of $\sim$ 50--70 degrees.
The antenna temperature $T_{\rm A}^{*}$  calibration 
was done by the standard chopper wheel method every 30 minutes to 1 hour.   
The data were converted to the main beam brightness temperature (\tmb) scale.  
The \hpbw\ at 691 GHz is $\sim$10\arcsec.
The main beam efficiency at the frequency of the CO $6-5$ transition at an elevation of 49\deg~ was estimated to be $\sim$ 40 $\%$,
through observations of Uranus.  
Pointing was checked approximately every two hours.
The pointing accuracy is estimated to be $\sim$ 5\arcsec.   
Daily gain variations as well as gain curve over different elevation angles were derived and corrected using CO $6-5$ spectrum of NGC 7027.  
Frequency calibration was done every 30 minutes.   
The data were reduced using the program CLASS of the GILDAS software. 

\notetoeditor{Please insert Table 1 here.}

\section{RESULTS AND ANALYSIS} \label{results}
\subsection{Submillimeter Dust  Continuum}\label{sharc}
\subsubsection{Overall Structures and Flux Measurements}\label{overalldust}
\begin{figure}
\epsscale{.70}
\includegraphics[angle=-90,scale=.9]{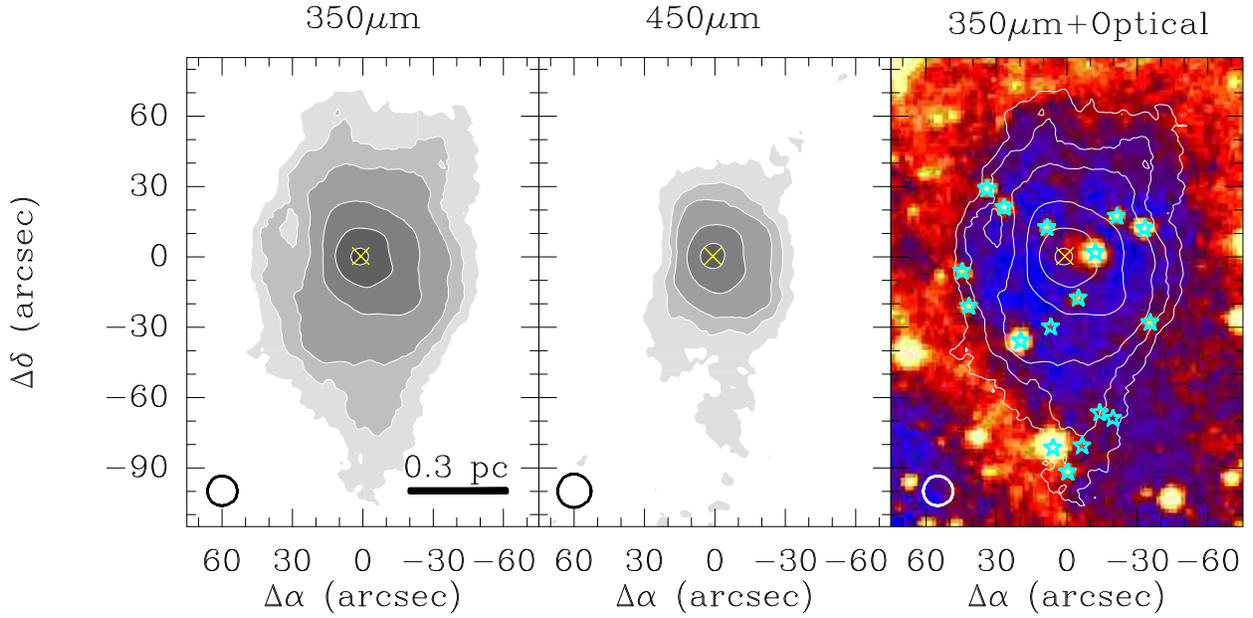} 
\caption{ SHARC II  350$\mu$m (left panel) 450$\mu$m (middle panel) images of 
the dense clump.  
The 350$\mu$m image is overlaid on the optical image (DSS 0.5$\mu$m, color) in the right panel.  
Blue stars mark the positions of foreground stars. 
Grey scale contours of two images from left are drawn at 2$\sigma$, 
5$\sigma$, 
9$\sigma$, 27$\sigma$ 81$\sigma$, and 243$\sigma$,
where 1$\sigma$=200~mJy beam$^{-1}$. 
The yellow cross at the center represents the position of the clump center, 
namely the peak of the 350$\mu$m continuum emission.  
The beams 
is shown in the left bottom corner of each panel.  
\label{sharcimage}}
\end{figure}

Submillimeter continuum observations over a 4 pc region confirmed that 
the object is an isolated massive dense clump (Figure \ref{sharcimage}).  
The submillimeter dust peak positions in both bands coincide with the 1mm and 3mm continuum peak position 
(RA = 20$^{h}$ 14$^{m}$ 26.$^{s}$03, Decl. = 41\deg 13\arcmin 32.\arcsec7 (J2000);   
 l = 78\deg 07\arcmin 20.\arcsec2, b = 03\deg 37\arcmin 58.\arcsec3)), measured by \citet{ces99} and \citet{she00}.   

The 350$\mu$m image traces 
a north-south elongated single dense clump with a size of $\sim$1.3 $\times$ 0.7 pc, 
which fits the silhouette against diffuse ionized emission.  
The 450$\mu$m dust continuum, which has a size of
about 
0.9 $\times$ 0.4 pc, looks more compact, with some extension towards the south.
The clump has a prolate spheroidal shape.  
The effective radius of the clump is defined as
$R_{\rm eff} = (A/\pi-\theta_{\rm HPBW}^2/4)^{1/2}$, 
where $A$ is the area of 2$\sigma$ contour traced by 350 $\mu$m and 450$\mu$m continuum.
We find $R_{\rm eff 350}=0.48$~pc and $R_{\rm eff 450}=0.37$~pc, corresponding to angular radii of 66\arcsec\ and 51\arcsec, respectively. 

\notetoeditor{Please insert Figures 1 here.}

The peak flux values at 350$\mu$m and 450$\mu$m are 
59.1$\pm$ 4.1 Jy beam$^{-1}$ and 27.4$\pm$1.2 Jy beam$^{-1}$, respectively.  
Within the uncertainties, these values are consistent with previous flux measurements done with SCUBA \citep{ces99}
with \hpbw=15\arcsec,
but the 350$\mu$m peak flux value is higher than 
another SHARC measurement at the same wavelength, 34 \jybm \citep{hun00},  with \hpbw = 11\arcsec.  
All peak flux densities at different wavelengths reported to date are summarized in Table \ref{tbl-2}.  
Our total flux densities computed inside the 2$\sigma$ contour are 477 $\pm$ 13 Jy  
at 350$\mu$m and 137 $\pm$ 14Jy at 450$\mu$m, respectively.  

\subsubsection{Radial Density Profile}\label{densityprofile}
\begin{figure}
\includegraphics[angle=-90,scale=.80]{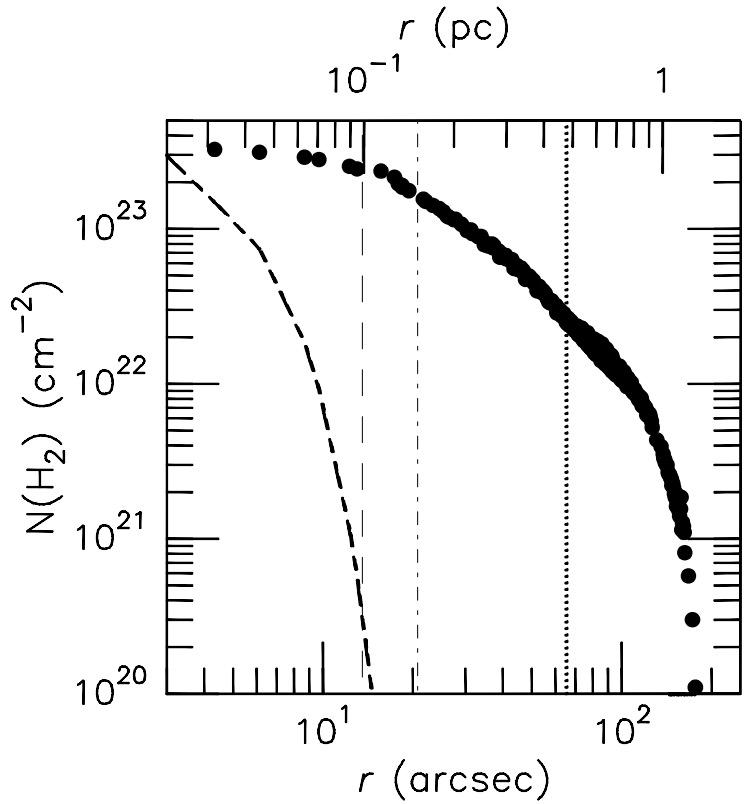} 
\caption{Radial column density profile of the dense clump (black dots) derived using the 350$\mu$m data.  
The thick broken line represents the beam shape.  
The vertical dotted line marks $R_{\rm eff 350}$. 
The dash-dotted and long-dashed lines mark the effective radii of the continuum emission 
measured at 1.2mm and at 850$\mu$m, respectively.  
 \label{columndensitydist}
}
\end{figure}

We approximate the
column density ($N$(H$_{2}$)) as a function of clump radius ($r$)  as a power law,
$N$(H$_{2}$)$\propto r^{p}$. This
reflects the density structure of the clump, which to a first order approximation
should also scale as a power law, with index $p-1$.  
Figure \ref{columndensitydist} represents the 
azimuthally averaged column density profile derived from the 350$\mu$m data.  
The density profile is plotted from the clump center, i.e. the peak position at 350$\mu$m.  
A discussion on the effect of a temperature gradient on the estimate of the
column density profile will be given in $\S$ \ref{dustprop}.  

One can see that
the inner part of the dense clump has a flatter index compared to that of outer part.  
The power law for $0.03 \la r \la 0.13$ pc 
(corresponding to 4.4\arcsec~ and 17\arcsec) 
is $r ^{-0.26 \pm 0.03}$.  
Note that this region is larger than \hpbw$_{350}$/2 = 4.\arcsec3, so that
the flatter index observed in the inner region is not due to beam dilution.
In the range $0.13 \la r \la 0.3$ pc 
(i.e. between 17 and 40\arcsec), the column density decreases as $r^{-1.32 \pm 0.02}$.  
Finally, for $0.3 \la r \la 0.7$ pc 
(40 -- 100\arcsec), the index $p$ 
is $-1.98 \pm 0.02$.  Steepening of the index in this region is likely due to
geometrical effects close to the clump border at $\sim$50\arcsec\ from the center.
Instead, in the two inner regions the estimates obtained are reliable and
imply
power law indices of $-1.26$ and $-2.32$ for the volume density.  
These issues will be further discussed in $\S$ \ref{dustprop}.

\notetoeditor{Please insert Figures 2 here.}

\subsubsection{350$\mu$m/450$\mu$m Flux Ratio Map}\label{sharcratiosec}
The dust continuum maps taken at two different wavelengths allow us to calculate a flux ratio map.
Note that such a ratio depends on two
parameters: the emissivity spectral index, $\beta$, and the temperature of the dust grains, \tdust.  

\begin{figure}
\epsscale{1}
\includegraphics[angle=-90,scale=.8]{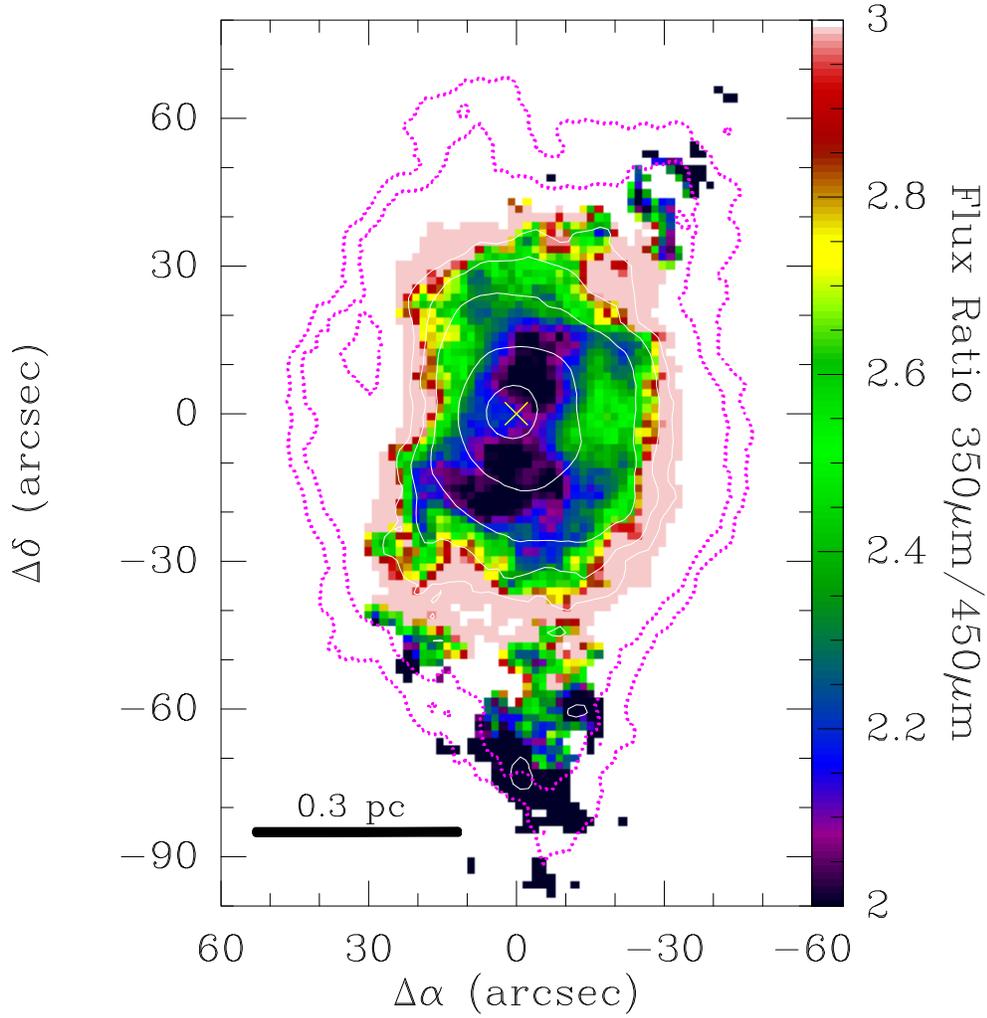} 
\caption{350$\mu$m/450$\mu$m flux ratio map. 
The 450$\mu$m emission is drawn in white contours corresponding to 3$\sigma$, 5$\sigma$, 9$\sigma$, 27$\sigma$, and 81$\sigma$. 
The pink dotted contours correspond to the 3$\sigma$ and 5$\sigma$ levels of the 350$\mu$m emission.  
The yellow cross 
marks the clump center. 
 \label{sharcratioonly}
}
\end{figure}

Figure \ref{sharcratioonly} shows a map of the 350$\mu$m/450$\mu$m flux ratio.
Note that the 350$\mu$m data have been smoothed to the same angular resolution as the 450$\mu$m image.  
Only data above 2 $\sigma$ 
have been used in the computation of the ratio.
The plotted ratio range is between 2 and 3 in order to show the change of the ratio clearly.  

The flux ratio map reveals a few intriguing features.
First of all, the flux ratio towards the clump center is low ($\sim$ 2)
and shows a bipolar structure elongated in the north-south direction.  
Second, the flux ratio increases towards the outer region and reaches the maximum ($\sim$ 3) at $r \sim 0.3$ pc.
Such a high flux ratio is found in all directions and traces a region elongated in the north-south direction.
Third, the flux ratio decreases down to $\lesssim$ 2 beyond the high flux ratio region at $r \gtrsim 0.4$ pc.  

\begin{figure}
\epsscale{1}
\includegraphics[angle=-90,scale=.6]{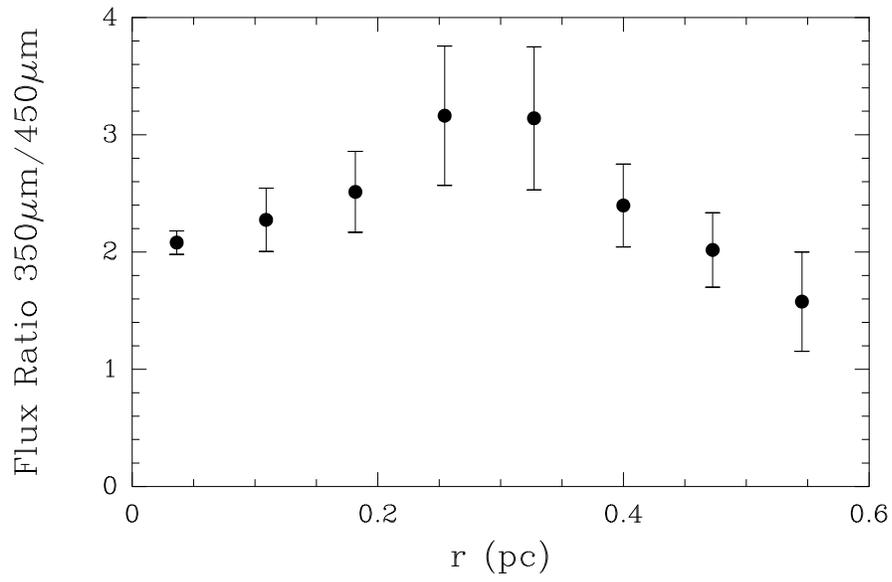}
\caption{Plot of the 350$\mu$m/450$\mu$m flux ratio as a function of the clump radius.
The bars are the standard deviation of the ratio in each annulus. 
 \label{sharcratio-radius}
}
\end{figure}

This variation of the flux ratio can be found also in Figure \ref{sharcratio-radius}.  
Here, 
the flux ratio is calculated over circular annuli with 10\arcsec~ width and 10\arcsec~ step. 
The error bars are the standard deviation of the ratio values measured in each annulus. 
Note that error of the flux ratio increases towards the clump edge, as the intensities of the continuum emission decrease.  

Considering the fact that there is only one prominent heating source, the massive (proto)star, inside the dense clump, 
the flux ratio pattern is expected to be spherically symmetry.  
However, the observed pattern of the flux ratio map looks significantly elongated, which
may be due to the bipolar outflow from the massive (proto)star.  
This will be discussed in detail in $\S$ \ref{discussion}.

\notetoeditor{Please insert Figures 3 and 4 here.}

\subsection{Molecular Components Traced with CO Emission}\label{co}
We use our maps of the clump in the CO $J=6-5$ line 
to investigate the physical properties of the molecular gas ($\S$ \ref{co765text}) as well as 
its low-velocity ($\S$ \ref{comaps}) and
high-velocity components ($\S$ \ref{co65outflows}).

\subsubsection{Physical Properties Estimated From CO $J=6-5$ and $J=7-6$ Spectra}\label{co765text} 
\begin{figure}
\epsscale{.60}
\includegraphics[angle=0,scale=.7]{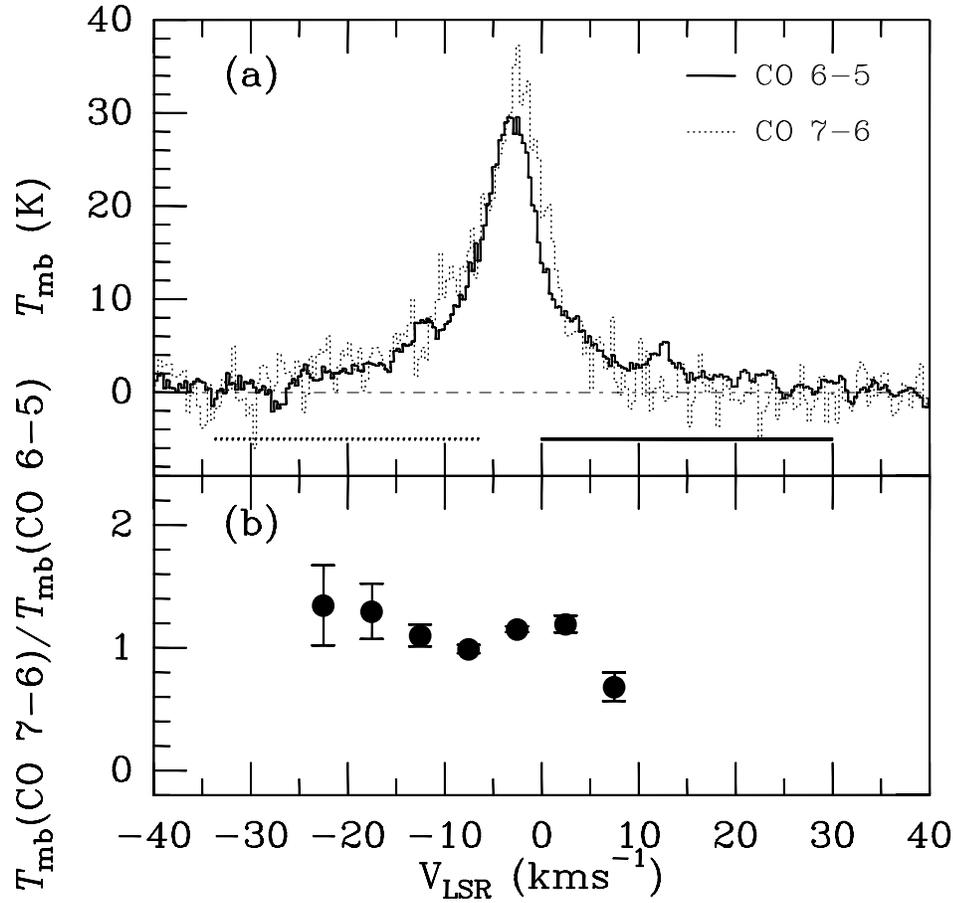} 
\caption{ (a) CO $6-5$ and $7-6$ spectra observed towards the center of the clump.  
The solid and dashed bars under the spectra indicate the ranges of 
red- and blue-shifted components (see $\S$ \ref{co65outflows}).  
(b)  Ratio of the brightness temperatures of the CO $7-6$ and $6-5$ emission.
\label{co765}
}
\end{figure}

In Figure \ref{co765} (a), the CO $J=7-6$ spectrum 
(Kawamura et al 1999) is compared with the CO $J=6-5$ spectrum to infer the physical properties of the molecular gas. 
Figure \ref{co765} (b) shows the ratio of the main beam temperatures of 
these transitions.  Since the temperature scale of the 
CO $7-6$ spectrum is a lower limit \citep{kaw99}, the brightness temperature ratio is to be taken as an upper limit.  
One can see that the red shifted wing has a higher ratio 
than the blue shifted one.  
\notetoeditor{Please insert Figures 5 here.}

When dust couples well with molecular gas, 
the excitation temperature of the gas must be close to the temperature estimated from 
the spectral energy distribution (SED) of the dust continuum.  
Therefore line emission is expected to be optically thin when \tmb~ is lower than \tdust=\tex\ ($\sim$ 40 K, Shepherd et al. 2000;  
see also $\S$ \ref{dustprop}).
Thus, the CO emission in the line wings is likely optically thin, because \tmb $\ll$ 40~K at high velocities.  

We now wish to obtain an estimate of the gas temperature in the high velocity gas (i.e. in the
line wings) using the result in Figure~\ref{co765}~(b).
For this purpose we express the ratio between the main beam 
brightness temperatures of the CO $7-6$ and $6-5$ lines as
\begin{equation}
 \frac{T_{\rm{mb~ CO}~ 7-6}}{T_{{\rm mb~ CO}~ 6-5}} = \frac{f_{7-6}}{f_{6-5}} \cdot \left(\frac{\nu_{6-5}}{\nu_{7-6}}\right)^2 \cdot
 \frac{B_{\nu_{7-6}}(T_{\rm ex})-B_{\nu_{7-6}}(T_{\rm bg})}{B_{\nu_{6-5}}(T_{\rm ex})-B_{\nu_{6-5}}(T_{\rm bg})} \cdot
 \frac{1-{\rm exp}(-\tau_{7-6})}{1-{\rm exp}(-\tau_{6-5})} 
\end{equation}
where
$T_{\rm{mb}}$ is the main beam brightness temperature of the CO transition, 
$f$ is the corresponding beam filling factor,
and $B_{\nu_{J-(J-1)}}(T) = \frac{2h \nu^{3}}{c^{2}}\cdot[{\rm exp}(\frac{h\nu_{J-(J-1)}}{kT})-1]^{-1}$. 
is the Planck function.
Here $h$ and $k$ are the Planck and Boltzmann's constants. 
$T_{\rm bg}$ is the cosmic microwave background temperature (2.7 K).  
At the frequencies of the CO $7-6$ and $6-5$ transitions 
($\nu_{7-6}$= 806.652~GHz; $\nu_{6-5}$= 691.473~GHz), 
$B_{\nu_{6-5}}(T_{\rm bg})$ and $B_{\nu_{7-6}}(T_{\rm bg})$ are negligible.  
As already pointed out, in the line wings $\tau\ll1$, so that [1-exp(-$\tau$)] $\simeq \tau$.  
Since the source is resolved, we also assume $\frac{f_{7-6}}{f_{6-5}} \simeq 1$. 
Under these approximations, the above equation can be re-written as 
\begin{equation}
\frac{T_{\rm{mb~ CO}~ 7-6}}{T_{{\rm mb~ CO}~ 6-5}} \simeq \frac{\nu_{7-6}}{\nu_{6-5}} \cdot
\frac{\exp\left(\frac{h\nu_{6-5}}{kT_{\rm ex}}\right)-1}{\exp\left(\frac{h\nu_{7-6}}{kT_{\rm ex}}\right)-1} \cdot
\frac{\tau_{7-6}}{\tau_{6-5}}
\end{equation}
Under local thermal equilibrium (LTE) condition, the ratio of the optical depths of the two transitions can be calculated from the following equation; 
\begin{equation}
\frac{\tau_{7-6}}{\tau_{6-5}} \simeq \frac{\nu_{7-6}}{\nu_{6-5}} \cdot \exp\left(-\frac{h\nu_{7-6}}{kT_{\rm ex}}\right) \cdot
 \frac{{\rm exp}(\frac{h\nu_{7-6}}{kT_{\rm ex}})-1}{{\rm exp}(\frac{h\nu_{6-5}}{kT_{\rm ex}})-1}
\end{equation}
Substituting Equation (3) in the Equation (2), replacing $h\nu_{7-6}/k=38.714$~K and $\nu_{7-6}/\nu_{6-5}=7/6$,
and solving with respect to $T_{\rm ex}$, one finds
\begin{equation}
T_{\rm ex} = \frac{38.714~{\rm K}}{\ln\left[\left(\frac{7}{6}\right)^2\frac{T_{\rm{mb~ CO}~ 6-5}}{T_{{\rm mb~ CO}~ 7-6}}\right]}  ~~~~. 
\end{equation}

Based on this equation, the blue- and red-shifted components have temperatures equal
to 130~K and 30~K respectively and we thus suggest that the temperature of the blue-shifted gas is
higher than that of the red-shifted gas.  
Please note that the difference on the ratio of the brightness temperatures of the CO lines could be partly 
due to difference in densities in the two outflow lobes.  

We also attempted an estimate of the optical depth of these transitions using the large velocity gradient (LVG) model.  
Assuming an 
H$_{2}$ volume density of 10$^{5}$$-$10$^{6}$~cm$^{-3}$, a CO column density of 10$^{22}$$-$10$^{23}$~cm$^{-2}$,  
an abundance ratio between molecular hydrogen and CO of 10$^{4}$ (see also $\S$ \ref{co65outflows}), 
and a measured line width of 25 \kms, 
we obtain $\tau_{7-6}=1.2$ and $\tau_{6-5}=0.7$ from the model.
The tau ratio is 0.7, similar to that obtainable from Equation~(3).

\subsubsection{Narrow Line-Width Rotating Massive Dense Clump}\label{comaps}
\begin{figure}
\epsscale{.60}
\includegraphics[angle=-90,scale=1]{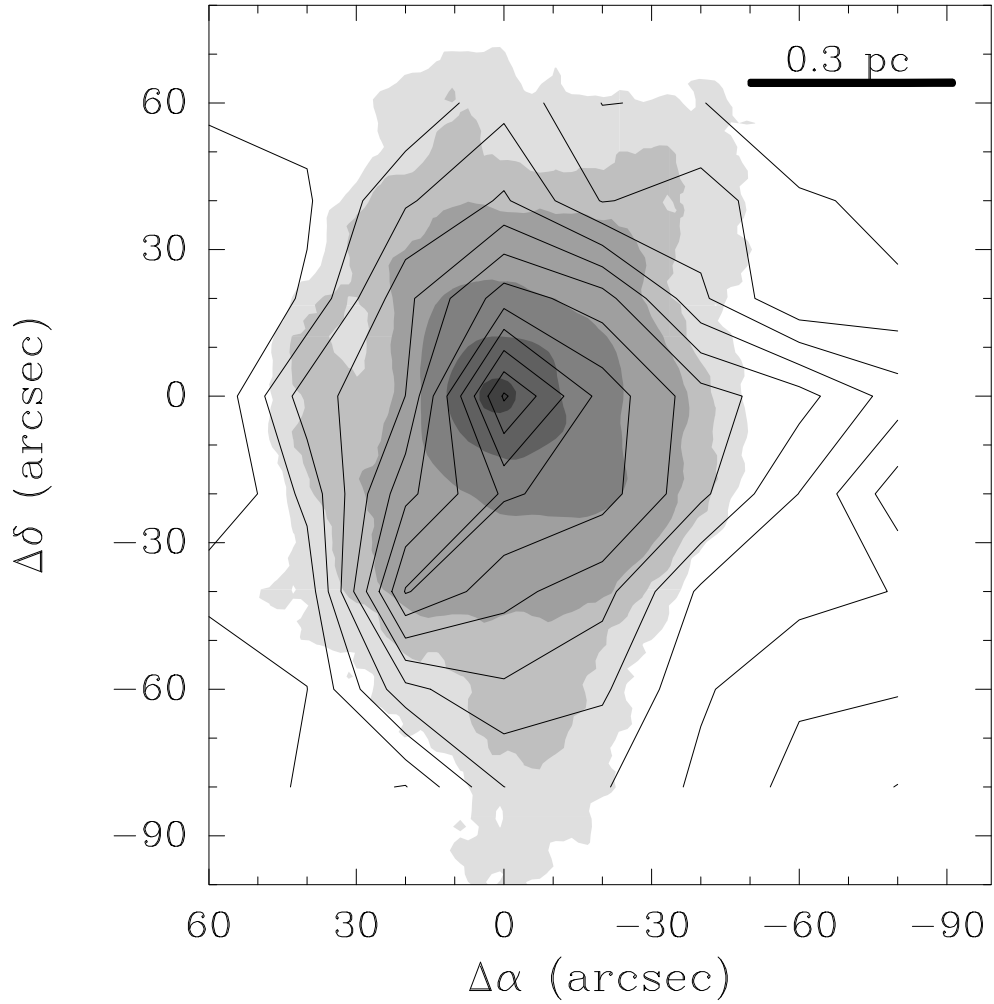} 
\caption{Comparison between CO $6-5$ line emission close to the systemic velocity (black solid line contours) 
and the 350$\mu$m dust continuum emission (grey scale contours, same as Figure \ref{sharcimage} left panel).   
Contours of the CO $6-5$ map are drawn in steps of 2$\sigma$, starting from the 2$\sigma$ level (10.6 K \kms).  
\label{co65quietgas}
}
\end{figure}

Figure \ref{co65quietgas} shows a comparison between the 350$\mu$m dust continuum and 
CO $6-5$ line emission with LSR velocity in the range from $-6.75$ to $-0.25$ \kms.  
This velocity range is chosen so that the emission in the range corresponds to the extended components.  
\notetoeditor{Please insert Figures 6 here.}

As shown in Figure \ref{co65quietgas}, the warm gas traced by the CO $6-5$ line shows good spatial correlation  
with the 350$\mu$m dust continuum emission. 
The molecular component traced by CO $6-5$ is not spherically symmetric.
The effective radius of the CO $6-5$ emission, $R_{\rm eff~CO 6-5}$, is 0.56 pc (77\arcsec), 
larger than $R_{\rm eff~350}$ 
($\S$ \ref{overalldust}).  
There is a strong CO $6-5$ peak towards the clump center (i.e. the continuum peak).  

The column density of the CO molecule using $J=6-5$ transition can be derived from 
(e.g., Equation (A1) of Scoville et al. 1986):
\begin{equation}
N_{\rm CO 6-5} = \frac{3k}{8\pi^{3}B_{\rm CO}\mu_{\rm CO}^{2}} \cdot  \frac{e^{\frac{E_{\rm CO _{J=5}}}{kT_{\rm ex}}}}{6}\cdot\frac{T_{\rm ex}+\frac{hB_{\rm CO}}{3k}}{1-e^{\frac{-h\nu_{\rm CO 6-5}}{kT_{\rm ex}}}}\cdot\frac{ \int T_{\rm mb} dv}{J_{\rm CO 6-5}-J_{\rm bg 6-5}}, 
\end{equation} 
where $B_{\rm CO}$ is the rotational constant of the molecule (57897.5 MHz), $\nu _{\rm CO 6-5}$ the frequency of the transition, 
$J(T)=\frac{\frac{h\nu_{\rm CO 6-5}}{k}}{exp(\frac{h\nu_{\rm CO 6-5}}{kT_{\rm ex}})-1}$ the specific intensity, and $v$ is the velocity.  Note that \Tmb = $\frac{\tau_{v}}{1-e^{-\tau_{v}}}T_{b}$, where $T_{\rm b}$ is the brightness temperature.  
For $\int T_{\rm mb} dv$, we use an integrated spectrum of the area inside the 2 sigma contour.  
Since CO $7-6$ data are available only for the central region and 
we don't have isotopic line data,  we assume that the column density 
of molecular hydrogen is proportional to the observed velocity integrated $^{12}$CO emission, $\int T_{\rm mb} dv$.  

The LTE mass of the clump, $M_{\rm LTE}$, is calculated as follows: 
\begin{equation}
M_{\rm LTE}= A \cdot N_{\rm CO 6-5} \cdot \frac{[\rm H_{2}]}{[\rm CO]} \cdot \mu_{\rm G} \cdot m_{\rm H_{2}} 
\end{equation} 
with $\frac{[\rm H_{2}]}{[\rm CO]}$ abundance ratio of molecular hydrogen to CO (10$^{4}$), 
$\mu_{\rm G}$ mean atomic weight of the gas (1.36),  and $m_{\rm H_{2}}$ mass the hydrogen molecule. 
The derived LTE mass is 390 \Msun~ for $T_{\rm ex}$=40K. 
Hereafter, 40 K will be used as the temperature of the clump.  
Please note that the optical depth of CO 6-5 line that traces the extended clump components 
towards limited central region is likely to be thick.  
Accordingly, the LTE mass derived above is lower limit. 

\begin{figure}
\epsscale{.60}
\includegraphics[angle=-90,scale=.7]{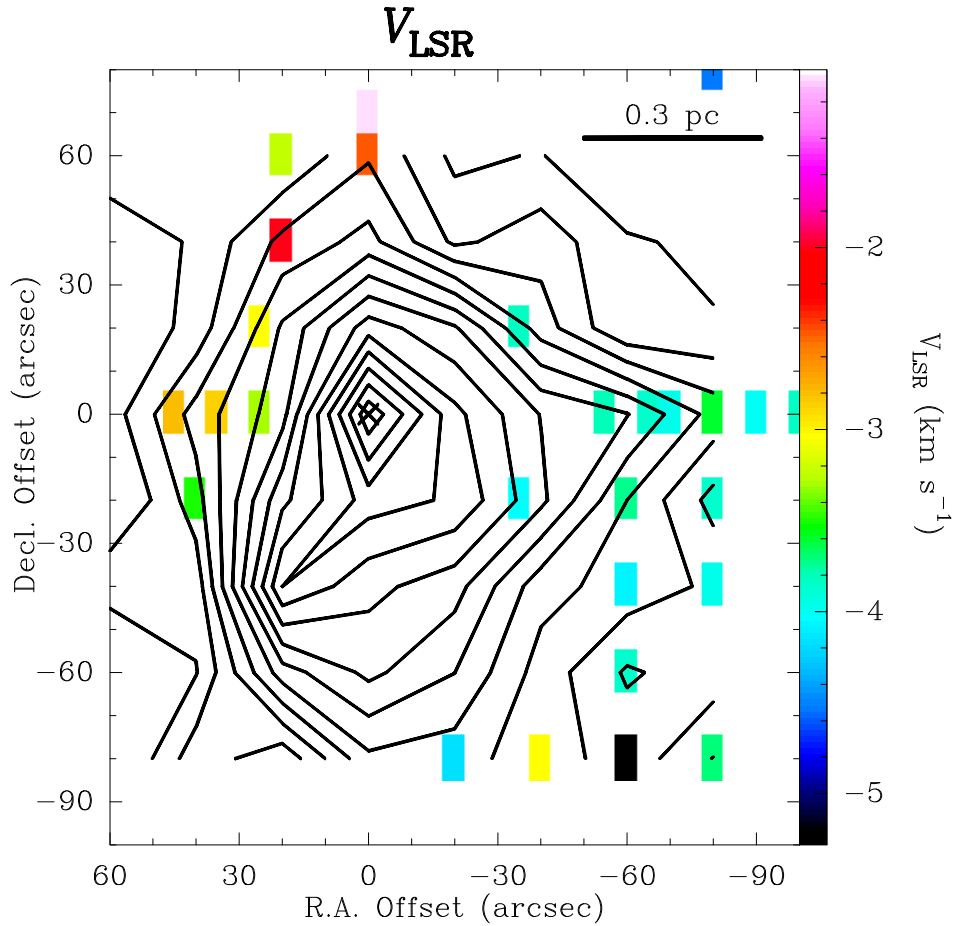} 
\caption{
The colored pixels in the diagram represent the LSR velocities of the CO $6-5$ narrow line-width components 
(contours, same as Figure \ref{co65quietgas}), overlaid 
on the map of the CO $6-5$ systemic velocity component (contours, same as Figure \ref{co65quietgas}).   
The black cross denotes the clump center. 
\label{co65quietgasvel}
}
\end{figure}

\begin{figure}
\epsscale{.60}
\includegraphics[angle=0,scale=.6]{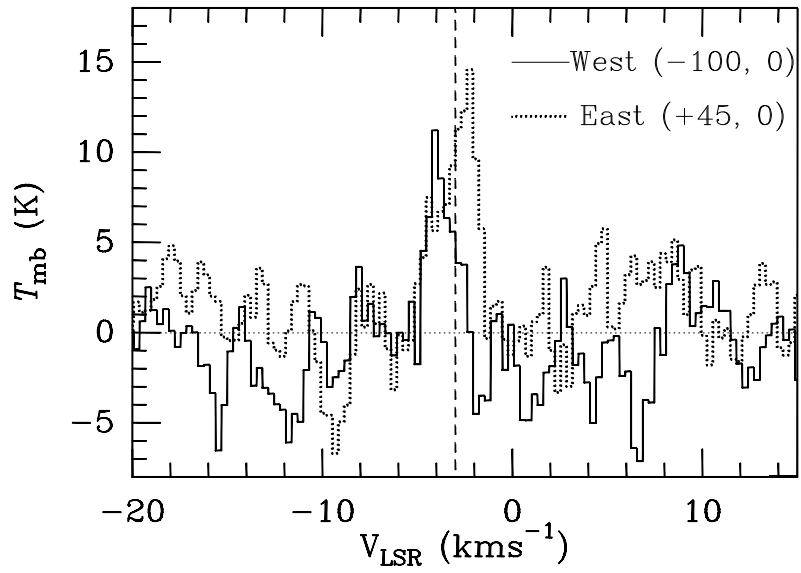} 
\caption{
CO $6-5$ spectra observed towards the positions at offsets (+45\arcsec, 0) (east) and (+100\arcsec, 0) (west) 
from the clump center. 
The dotted vertical line marks \vsix.  
\label{co65quietgasvelspec}
}
\end{figure}

\notetoeditor{Please insert Figures 7, 8 here.}
Figure \ref{co65quietgasvel} shows 
the LSR velocities towards selected positions where the full width at half
maximum (FWHM) of the line is $\le$ 3 \kms.
While the east part of the clump has LSR velocities ranging between $\sim -1$ to $-3.5$ \kms, 
in the west side the velocity ranges from $-3.0$ to $-5.2$ \kms.  
The velocity gradient measured over 1~pc is $\sim$ 2 \kms pc$^{-1}$.
The rotation angular velocity, $\Omega$, is $\sim 7\times 10^{-14}$ s$^{-1}$.  
In Figure \ref{co65quietgasvelspec} we show template spectra of the narrow line-width components 
observed in the east side and west side of the clump.  
This comparison demonstrates the clear velocity shift from east to west observed across the clump.  
Closely looking at the \vlsr change, the velocity shift might be from northeast to southwest.  
The direction of the velocity gradient is almost perpendicular to the major axis (directed north-south) of the massive dense clump.  
Assuming that the velocity shift is caused by rotation of the clump and that centrifugal forces balance gravitational ones,
one can estimate the equilibrium mass of the molecular dense clump as
$M_{\rm rot} \simeq \frac{\Omega^{2} \cdot r^{3}}{G}=200$~\Msun, 
where $G$ is the gravitational constant.  
This value agrees reasonably well with the LTE mass, supporting the idea that the velocity gradient 
may be largely due to rotation of the clump.  

We note that rotation of the narrow line-width molecular gas component occurs in the opposite sense
with respect to rotation of the circumstellar Keplerian disk associated with 
the (proto)star.
A similar discrepancy in the sense of rotation on different spatial scales,
has been observed also in W3 \citep{hay89} 
and Orion \citep{har83}, for massive star forming regions, and in
a starless core, L1521F (MC27) 
in the Taurus Molecular Cloud \citep{sh04a}, for low-mass star forming regions.
This suggests that such a complex kinematical structure may not be unusual  
in the star formation process.

\notetoeditor{Please insert Figures 9 and 10 here.}
The measured \dvfwhm~ averaged over the entire CO $6-5$ emitting region is 3.9 $\pm$ 0.2 \kms.  
 From the line width, one can derive the virial mass, \Mvir, of the clump:
\begin{equation}
M_{\rm VIR}\sim \frac{5+2e}{3+e} \cdot \frac{R_{\rm eff~CO 6-5}}{G} \sigma_{v_{\rm total}}^{2}  ~~~~~~~~~. 
\end{equation} 
where the volume density 
is proportional to r$^{-e}$.  
The virial mass 
is estimated to be 400\Msun for this dense clump. 
One can assess the equilibrium of the dense clump by comparing this value with the LTE 
mass and the one obtained assuming centrifugal equilibrium.
We note that $M_{\rm LTE} \simeq M_{\rm rot}$ and at the same time, 
$M_{\rm LTE} \simeq M_{\rm VIR}$. 
This suggests that the clump is in an equilibrium state and 
rotation of the core contributes significantly to the equilibrium of the clump.  

\begin{figure}
\epsscale{.60}
\includegraphics[angle=-90,scale=0.7]{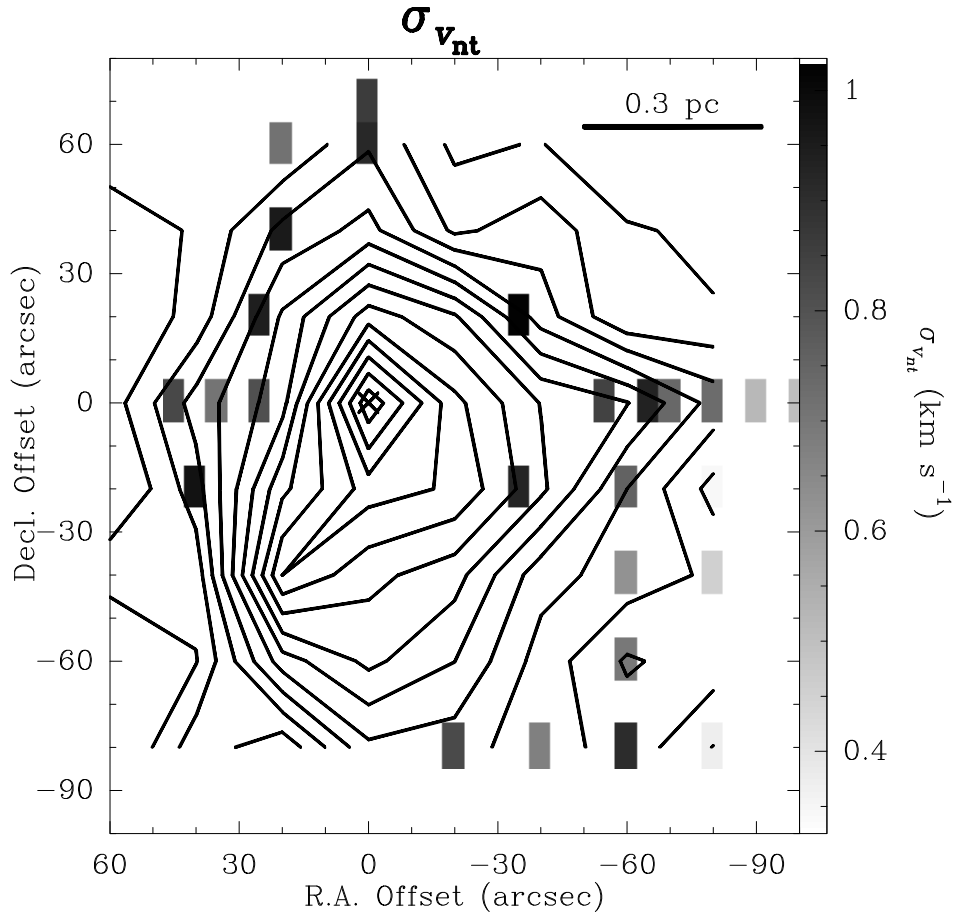} 
\caption{
The grey scale represents the non-thermal velocity dispersion ($\sigma_{v_{nt}}$) 
of the narrow line-width components, overlaid on 
the map of CO $6-5$ systemic velocity component (contours, same as Figure \ref{co65quietgas}).  
The black cross marks the clump center.  
\label{co65quietgasvelwidmap}
}
\end{figure}

\begin{figure}
\epsscale{.60}
\includegraphics[angle=0,scale=0.6]{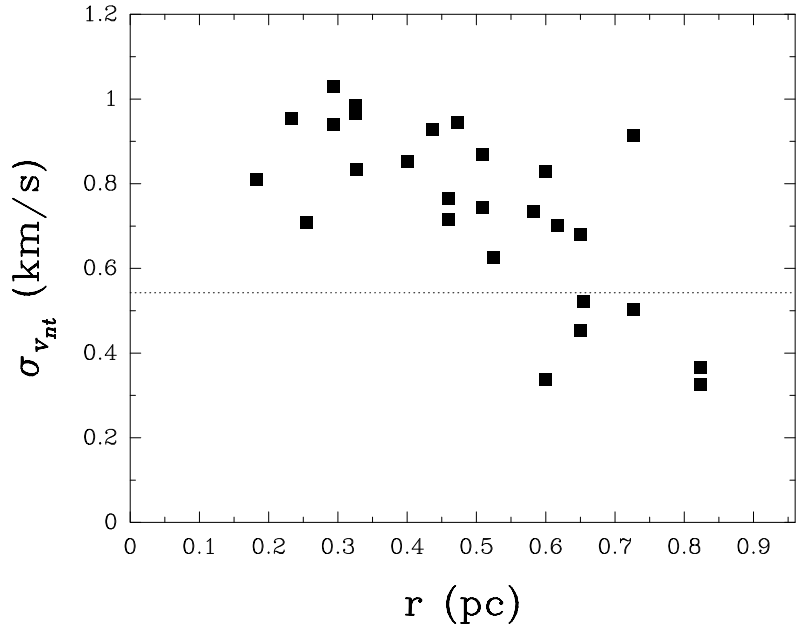}
\caption{
Plot of non-thermal velocity dispersion versus distance from the clump center.
\label{co65quietgasvelwid}
}
\end{figure}

Figure \ref{co65quietgasvelwidmap}  shows the non-thermal velocity dispersion, $\sigma_{v_{nt}}$, 
of the narrow line-width molecular gas components.  
Note that the diagram of the velocity width is plotted under an assumption of constant temperature over the core.  
Since the optical depth of the CO $6-5$ line is relatively small (see $\S$ \ref{co765text}), one can estimate the non-thermal 
velocity dispersion ($\sigma_{v_{nt}}$) by subtracting the thermal line width  
$(\frac{k T_{\rm k}}{m_{avr}})^{1/2}$ from total velocity dispersion, 
$\sigma_{v_{total}} = [\sigma_{v_{\rm obs}}^{2} - k T_{\rm k} (\frac{1}{m_{\rm CO}}-\frac{1}{m_{avr}})]^{1/2}$,  
where $\sigma_{v_{\rm obs}}$ is the observed velocity dispersion, 
$m_{avr}$ is the average mass unit, 
and $m_{\rm CO}=28$ amu is the mass of the CO molecule.
$\sigma_{v_{\rm obs}}$ can be expressed as $\Delta v_{\rm FWHM} (4~{\rm ln} 2)^{-1/2}$, 
where $\Delta v_{\rm FWHM}$ is the observed line FWHM.
 From Figure \ref{co65quietgasvelwid},
one can see that the non-thermal velocity dispersion decreases with increasing distance from the clump center.
To derive the non-thermal velocity dispersion, 40 K has been used as gas kinetic temperature.
The dotted line in the diagram shows the thermal velocity dispersion when \Tk = 40 K.  
One can see that the thermal velocity dispersion becomes dominant at $r$= 0.6 pc ($\sim$ 80\arcsec).  

\notetoeditor{Please insert Figures 11 here.}
\subsubsection{CO $6-5$ Large Scale Bipolar Outflow}\label{co65outflows}
\begin{figure}
\epsscale{.70}
\includegraphics[angle=0,scale=.9]{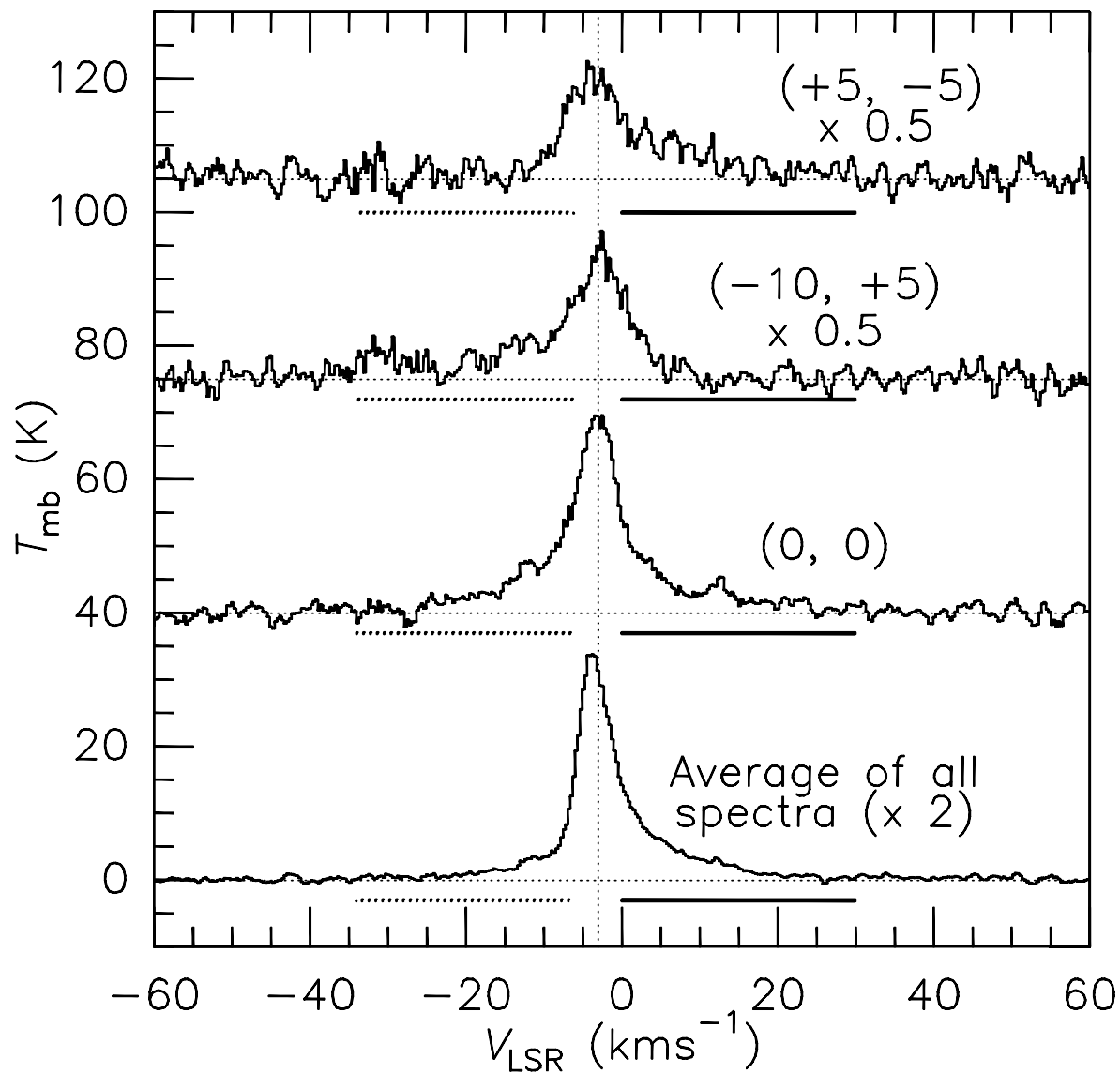} 
\caption{CO $J=6-5$ spectra taken towards three different positions 
((RA$_{\rm offset}$\arcsec, Dec$_{\rm offset}$\arcsec)=(+5, $-5$), ($-10$, +5), and (0, 0))
and average spectrum of all spectra taken within a central area of 1\arcmin $\times$ 1\arcmin 20\arcsec. 
The solid and dashed bars under each spectrum show the ranges of 
red- and blue-shifted components.  
The vertical dashed line is drawn at \vsix. \label{co65spectra}
}
\end{figure}

Figure \ref{co65spectra} shows the CO $J=6-5$ spectra taken towards three positions and the 
mean spectrum over a $\leq$ 120\arcsec~ area around
the clump center.   
One can see prominent wings in the spectra taken towards each outflow lobe.  
Blue- and red-shifted emission is detected over the velocity ranges $-34.0 \leq$ \vlsr ~\kms $\leq -6.0$ and 
0.0 $\leq$ \vlsr ~\kms $\leq$ +30, respectively.  
The spectrum observed towards the clump center (0\arcsec, 0\arcsec)  and the mean spectrum clearly show
both red- and blue-shifted emission.
The spectrum taken towards the center position has a strong peak at \vlsr=$-3$ \kms.  

\begin{figure}
\epsscale{.70}
\includegraphics[angle=-90,scale=.7]{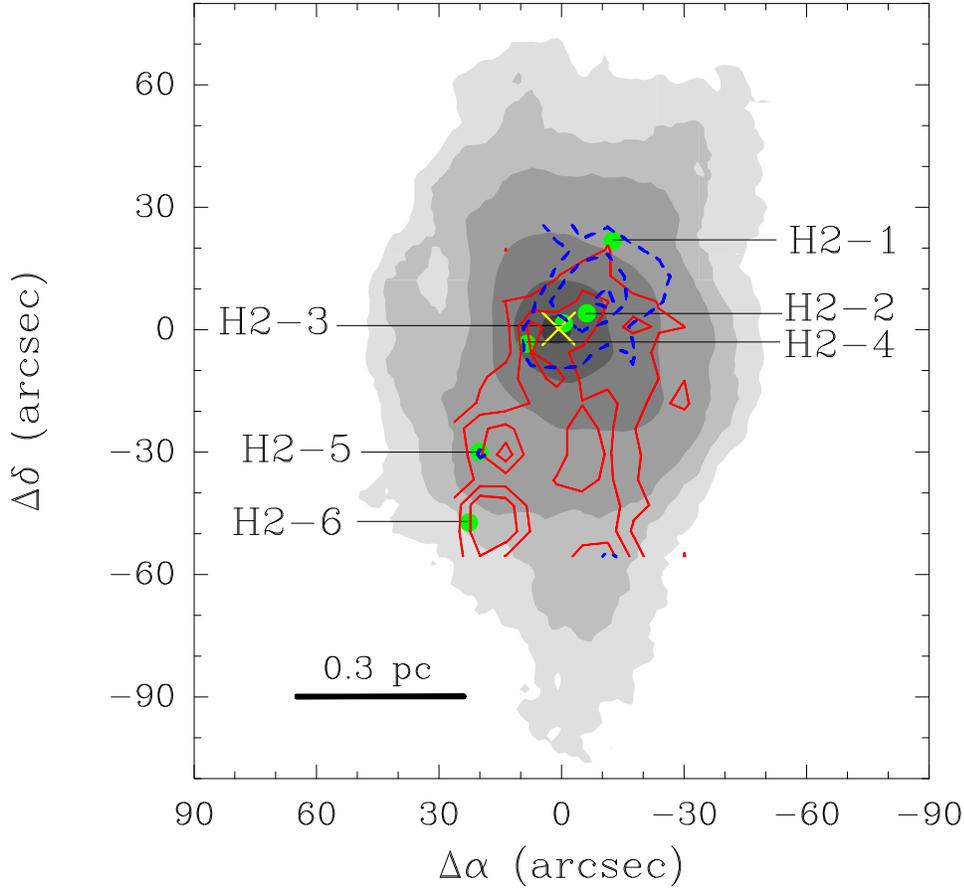} 
\caption{CO $J=6-5$ blue-shifted ($-34.0 \leq V_{\rm LSR}$ \kms $\leq -6.0$, 
blue dashed contours) and red-shifted (0.0 $\leq$ $V_{\rm LSR}$ \kms $\leq +30$, 
red solid contours) emission superposed on the 350$\mu$m image (same as Figure 1).  
Contours of the CO $J=6-5$ components are drawn in steps of 3$\sigma$, starting from the 3$\sigma$ level (50 K \kms).
The yellow cross marks the clump center.  
The shocked H$_{2}$ knots (green dots) marked and numbered in the diagram are taken from
\citet{she00}.  
\label{co65outflow}
}
\end{figure}

\notetoeditor{Please insert Figures 12 here.}
Figure \ref{co65outflow} shows the distribution of CO $J=6-5$ blue-shifted 
and red-shifted 
emission superposed on the 350$\mu$m dust continuum emission map.   
Both outflow lobes are located inside the 350$\mu$m dust continuum component.  
The red-shifted bipolar outflow lobe is more than twice extended compared with the blue-shifted outflow lobe.  
Integrated intensities of the blue- and red-shifted outflow lobes are measured to be 
$(9.6 \pm 0.9) \times 10^{2}$ and $\gtrsim (2.5 \pm 0.1) \times 10^{3}$ K \kms, respectively.  
The masses of the blue and red shifted outflows are estimated to be 28 and 13 \Msun, respectively.  
 From the outflow velocity, $\sim$30 \kms, and the extension of the outflow, $\sim$0.6 pc, one can 
estimate the dynamical time scale of the bipolar outflow.  The averaged dynamical time scale of the outflow is estimated to be  
$\gtrsim 1.7 \times 10^{4}$ years.  
Note that the estimated age is lower limit  
because the south edge of the red shifted outflow lobe isn't fully covered and 
the inclination angle is unknown.  The inclination angle affects the size and velocity.  
The outflow parameters derived from the CO $6-5$ and $1-0$ lines are summarized in Table \ref{tbl-3}.  
The outflow velocities obtained from the $6-5$ line are higher than those from the $1-0$, while
the mean dynamical time scale of the $6-5$ flow is shorter than that of the $1-0$.  
A more detailed discussion of the outflow properties will be given in $\S$ \ref{discussion}.

\section{DISCUSSION}\label{discussion}
\subsection{Dust Property and Column Density Profile} \label{dustprop} 
To derive the dust mass \Mdust,  
we have used the expression
(see Hildebrand 1983), 
\begin{equation}
M_{\rm dust} = \frac{F_{\nu} D^{2}}{B_{\nu}(T_{\rm dust})} \cdot\frac{1}{\kappa_{\nu}}, \label{emd} 
\end{equation} 
where $F_{\nu}$ is the measured flux density, 
$D$ the distance to the object, and $B_{\nu}$(\tdust) the Planck function.
The coefficient 1/$\kappa_{\nu}$ ($\kappa_{\nu}$ is dust opacity), 
is estimated to be 0.1 g$\cdot$ cm$^{-2}$ at 250$\mu$m in NGC7023  \citep{hil83}.  
$\kappa_{\nu}$ depends on the frequency as
$\kappa_{\nu} \varpropto \nu^{\beta}$. 
$\beta$ may range from $\sim$ 1  at $\lambda \lesssim 200 \mu$m to $\gtrsim$ 2 at $\sim$1 mm 
(Hildebrand 1983 and the references therein).
Extrapolating $\kappa_{\nu}$ from 250$\mu$m to 350$\mu$m, 
one obtains $\kappa_{850 GHz} \simeq 5.6$ cm$^{2}$ g$^{-1}$.
$\beta$ can be estimated from the flux ratio of 350$\mu$m and 450$\mu$m using the equation
\begin{equation}
\frac{F_{\nu_1}}{F_{\nu_2}}\cdot\frac{\Omega_{\nu_1}}{\Omega_{\nu_2}}=
\frac{B_{\nu_2}(T_{dust})}{B_{\nu_1}(T_{dust})}\cdot \left( \frac{\nu_1}{\nu_2} \right) ^{\beta}~~~,    \label{ebe}
\end{equation} 
where $\Omega_{\nu}$ is the beam solid angle.  
Note that in our case $\frac{\Omega_{\nu_1}}{\Omega_{\nu_2}}\simeq 1$,
because the 350$\mu$m image has been smoothed to the same angular resolution as the 450$\mu$m image.   

Assuming that the clump mass equals $M_{\rm rot} \simeq 200$~\Msun,
one may use Equation (\ref{emd}) to derive an estimate of the mean
clump temperature, $T_{\rm dust}=40$~K.
Then Equation (\ref{ebe}) gives $\beta\simeq1.7$.
This value lies within the typical range
$1.5 \lesssim \beta \lesssim 2.4$ for high mass star forming regions  
(e.g., Molinari et al. 2000).  
The dust temperature is more or less consistent with the temperature, $\sim$ 44~K, derived from the 
SED fitting by \citet{she00}, but 
lower than the 60~K estimated by \citet{ces99}. 
Note that the errors on $\beta$ and especially \tdust~ may 
be significant, as \mrot~ (200 \Msun) might not represent the true clump mass.  
\citet{hof07} estimates a clump mass of $\sim$ 400 \Msun from their SED fitting.  
The column density towards the clump center is estimated to be $\sim 3.4 \times 10^{23}$ cm$^{-2}$, using 
the values of $\beta$ and \tdust~ described above. 
This implies a volume density of $\sim10^{6}$ cm$^{-3}$ towards the clump center.  
Towards the outflow lobes, the estimated number density becomes $\sim 10^{5}$ cm$^{-3}$.  

A temperature gradient is likely to exist inside the clump, as this is
internally heated by the embedded (proto)star. Such a gradient is likely
to be similar to a power law of the type $T(r) \propto r^{-2/(4+\beta)}$ (Doty \& Leung 1994), 
with $r$ distance from the clump center. In our case $\beta=1.7$ and hence the
the power law index is $-0.35$.  
The effect of this gradient on the estimated density slopes inside the clump
is to change the power law indices of the inner ($0.13 \la r \la 0.3$ pc)
and outer ($0.3 \la r \la 0.7$ pc) regions to $\sim -0.9$ and $\sim -2$, respectively.
This has been assessed by recalculating the column density profile assuming 
the clump is spherically symmetric.  
These indices are not significantly different from 
those derived without taking the temperature gradient into account.  
For the non-thermal velocity dispersion described in $\S$ \ref{comaps}, 
the velocity dispersion won't change much (order of 0.01 \kms) even when 
we consider an temperature gradient of $T$($r$)$\propto r^{-0.35}$. 

An index of $-1.26$ for the $0.03 \la r \la 0.13$ pc region is only slightly shallower
than that typical of a region undergoing free-fall
($\rho(r) \propto r^{-\frac{3}{2}}$) -- as in the runaway collapse model (Larson-Penston-Hunter solution, 
Larson 1969; Penston 1969; Hunter 1977) 
and the inside-out collapse model by \citet{shu87}.
Instead, an index of $-2.32$ for the $0.13 \la r \la 0.3$ pc region
resembles the density profile $\rho(r) \propto r^{-2}$ 
of a system in hydrostatic equilibrium.
We thus speculate that
the inner region of the clump is experiencing infall, whereas the ``infalling wave'' has
not yet reached the outer region beyond $r \gtrsim 0.1$ pc. 

\subsection{Internal Structures of The Dense Clump}\label{sharcratiodiscuss} 
\subsubsection{CO $6-5$ Outflow and Shocked H$_{2}$ Knots}  \label{co65outflowsd}
 From Figure \ref{co65outflow}, 
one can see a spatial coincidence between the outflow lobes and the shocked H$_{2}$ knots.  
All six knots lie inside the CO $6-5$ bipolar outflow lobes.  
More specifically,
the CO $6-5$ emission shows clear peaks close to the H2-5 and H2-6 knots located in the red shifted lobe.
Such a correlation between CO $6-5$ peaks and shocked H$_{2}$ knots indicates that they are physically associated.  
Note that, unlike the CO $6-5$ emission, no peak towards the knot positions is found in the CO $1-0$ line, as one
can see in  Figure 1 of Shepherd et al. (2000). A likely interpretation is that
CO $6-5$ traces more excited gas compared to the $1-0$ transition.  
This might be due to the gas at the knot locations being heated and compressed by interaction
with the jet powered by the (proto)star.  

\subsubsection{Implications from the Submillimeter Dust Radio Map}  \label{sharcratio}  
There are three major components in the submillimeter flux ratio map (see $\S$ \ref{sharcratiosec} and Figure \ref{sharcratioonly}), namely:
(1) a bipolar feature in the central region of the clump, with flux ratio $\simeq 2 - $ 2.2;
(2) a region $\sim$ 0.5 pc in diameter and elongated north-south, with flux ratio starting from $\sim 2 - 3$ and 
increasing upto $\gtrsim$ 3 close to the outer border;
and (3) region of lower flux ratio, $\lesssim$ 2, outside of component (2).  
Component (1) is located inside component (2).  
The $PA$ of the bipolar structure is about $-17$\degree.
Note that features (1) and (2) are detected above a 3$\sigma$ level in the 450 $\mu$m map and above
a 5$\sigma$ level in the 350 $\mu$m map. 
Note that 450$\mu$m emission in component (3) is detected at a 2 $-$ 3$\sigma$ level,
whereas the 350$\mu$m emission is much stronger.
Therefore, we believe that all the three features represent real structures.  

Equation (\ref{ebe}) shows that the ratio between the 350 and 450$\mu$m fluxes
is an increasing function of both $\beta$ and (albeit more weakly) $T_{\rm dust}$. It
is thus impossible to decide whether a change in the ratio is due to a variation
of $\beta$, temperature, or both.
For example, for T = 40 K, a $\beta$ of 1 and 2 
implies a flux ratio of 1.9 and 2.4, 
respectively.  
Vice versa, for $\beta$=1.7, a temperature of 20, 40, and 60 K implies a flux ratio of 1.9, 2.3, and 2.4, respectively. 
 From these numbers one can see that, in order to reproduce flux ratios as
large ($\sim$3) as those measured on the surface of component (2), very likely an increase of both $\beta$ and $T_{\rm dust}$
is needed.
Hence component (2), which 
encases the bipolar feature, might be a cocoon of dense material with warm surface.  
On the other hand, the lower flux ratio value of component (3) indicates that
\tdust~ and/or $\beta$ in the region ahead of the outflow lobes are lower compared to component (2).  
We speculate that the higher \tdust\ and $\beta$ 
observed on the surface of the cocoon represented by component (2)
might be due to a weak shock, propagating from the (proto)star and possibly created by the outflow.
Considering a sound speed of $\sim$ 0.8 \kms, 
the crossing time of the weak shock would be $\la  3 \times 10^{5}$ years.
Note that this is likely an upper limit to the age of the outflow (and perhaps
of the (proto)star),
as shock waves may propagate with larger velocities than the sound speed.   

How to explain the low flux ratio in component (1)? The fact that
the shape of this region matches very well the outflow lobes
suggests a connection between the two. This suggests that the
flow might have swept away the smallest grains from that region,
thus biasing the grain size towards the largest ones.
This phenomenon would have the effect of reducing the
slope of the spectral index $\beta$ and hence of the flux ratio.

In conclusion, the flux ratio variations across the clump strongly suggest that the massive dense clump is no longer uniform 
in terms of temperature as well as dust properties.   
To construct a model to explain the observed features, one must keep in mind
such inhomogeneities.  

\subsubsection{Comparison Between Submillimeter Flux Ratio Map And Molecular Components}\label{sharcratio-outflow} 
\begin{figure}
\epsscale{.5}
\includegraphics[angle=-90,scale=.9]{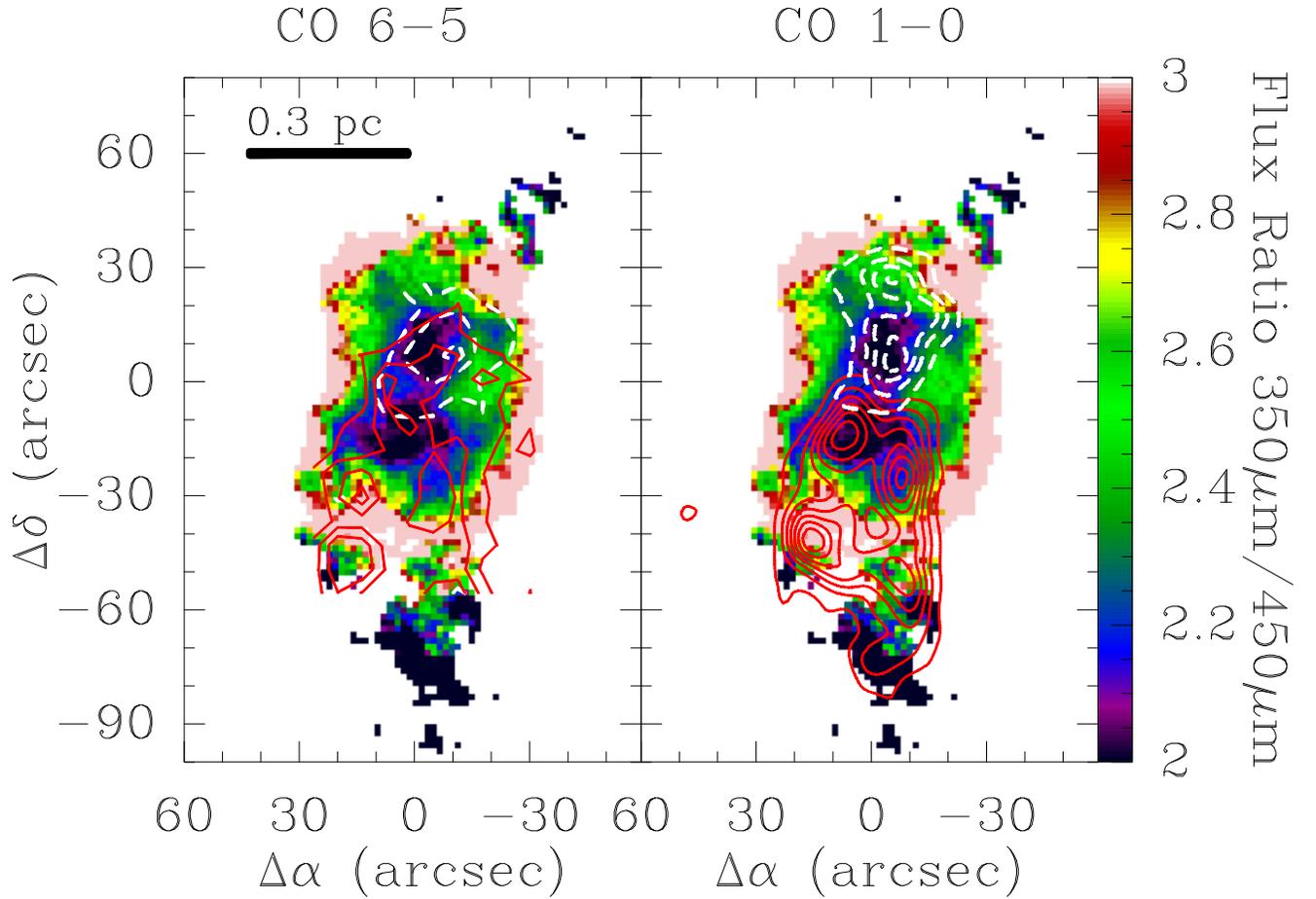} 
\caption{
CO $6-5$ blue- and red-shifted (left, same as Figure \ref{co65outflow}), 
and CO $1-0$ blue- and red-shifted (right, same as Figure 1 of Shepherd et al. 2000) maps, 
superposed on the 350/450$\mu$m flux ratio map.
White dashed and red solid contours 
denote the blue- and red-shifted lobes respectively.
\label{sharcratiooutflow}
}
\end{figure}

Here we make a detailed comparison between the flux ratio map and the structure of the dense clump and outflow.
Figure \ref{sharcratiooutflow} shows the CO $6-5$ and CO $1-0$ outflow lobes superposed on the flux ratio map.  
We have already discussed the good correlation between the CO outflow and the bipolar component (1) of the flux ratio map.
Now, we note also that
the peak of the blue shifted CO 6--5 lobe is located on the west edge of the northern bipolar feature, which
may be due to interaction with the jet detected in SiO and H$_2$ line emission (Cesaroni et al. 1999)
and oriented with PA$\simeq-60$\degree.
This jet might have excited the molecular gas lying on the west side of the northern lobe.  
Nothing similar is seen to the SW in correspondence to the CO $6-5$ red shifted outflow lobe.
There are two distinct CO $6-5$ red shifted peaks associated with the shocked H$_2$ knots, H2-5 and -6, 
located at (14\arcsec, -30\arcsec) and (17\arcsec, -48\arcsec) respectively, as described in $\S$ \ref{co65outflowsd}. 
At the position of the H2-5 knot, a high value of the flux ratio ($\sim$ 3) is observed, \
suggesting relatively large values for both the temperature and $\beta$.
In correspondence to the CO $6-5$ peak associated with knot H2-6, the flux ratio is $\lesssim$ 2.5.  
The difference in flux ratio between H2-5 and H2-6 may indicate
that the temperature and/or $\beta$ at the position of knot H2-6 could be lower than those of H2-5.

\section{SUMMARY}\label{remarks}
An observational study of the IRAS 20126+4104 clump 
is presented.  The main findings of this study are summarized as follows.  

(I) The 350$\mu$m and 450$\mu$m dust continuum images revealed 
that 
the power law index of the volume density in the inner part of the clump is $-1.26 \pm 0.03$ ($r \leq 0.1$ pc), 
shallower than that in the outer part, $-2.32 \pm 0.02$ ($0.3 \leq r \leq 0.1$ pc).  
The power law indices 
suggest that the inner region might be infalling, while the outer region of the clump is still in hydrostatic 
equilibrium.    


(II) A map of the peak velocity of narrow line-width CO 6--5 components reveals a velocity
gradient across the clump, suggesting rotation about an axis oriented  northwest$-$southeast. The rotation
occurs in the opposite sense with respect to the circumstellar Keplerian disk associated
with the massive (proto)star. 

(III) Assuming centrifugal equilibrium, one estimates an enclosed mass \Mrot~ of $\sim$ 200\Msun.  
Estimated \Mlte~ and \Mvir~ ($\sim$ 400\Msun) of the clump are comparable to \Mrot.  
This indicates that the clump is in equilibrium state as a whole and rotation contributes a large fraction 
to support the clump from gravitational collapse on large clump scale.  
At the radius of 0.1 pc, the rotation may be disconnected from the inner region and 
the medium inside of $r \sim$ 0.1 pc may be infalling.  
This might be partly related to the disagreement of the rotational axes of the large scale 
clump and of the disk associated with the massive (proto)star.  
%


(IV) 
The mean spectral index $\beta$ and dust temperature are estimated  to be 
1.7 and 40~K, respectively.  


(V) The 350$\mu$m/450$\mu$m ratio map outlines three major components, namely:
(1) a bipolar feature, which may represent the region swept by the bipolar outflow lobes;
(2) a dense cocoon $\sim$ 0.5 pc in diameter, with a warm surface; and 
(3) a low temperature/low $\beta$ component outside component (2).
These three components indicate that 
the temperature as well as the nature of the dust (grain size distribution and composition) may vary significantly 
inside the dense clump.  
The warm layer of component (2) might have been created by a shock propagating from the 
(proto)star.  
The crossing time of the shock is estimated to be $\gtrsim$ 3 $\times$ 10$^{5}$ years. 

(VI) Strong correlation between CO $6-5$ peaks and two of the shocked H$_{2}$ knots are identified, 
indicating they are physically associated.



\acknowledgments
This research was performed at the Caltech Submillimeter Observatory, supported by 
NSF grant AST-0229008.  
HS is grateful to Hiroshige Yoshida for the observational support in July 2004 
and Debra Shepherd and J. H. Kawamura for providing us their 
data. 
HS is indebt to Dr. Peter Schilke for his LVG code.   
Thanks are also due to Thomas A. Bell, John Carpenter and Roger Hildebrand   
for valuable comments and discussions.  
HS also thank the Gildas development team 
(http://www.iram.fr/IRAMFR/GILDAS) which provides us their  
data reduction packages.  
We thank the Hawaiian people for allowing us to observe from the summit of their sacred mountain, 
Maunakea.  



Facilities: \facility{CSO} 



\appendix
\section{Appendix: Distance To The Massive Dense Clump}
IRAS 20126+4104 is located in the northwest side of the Cygnus X region.  
This region \citep{pid52,dav57} is one of the most active star forming complexes in the Galaxy.   
Here one can find a large variety of objects, including OB associations,  
atomic diffuse clouds, molecular dense clouds/cores, evolved stars, and supernova remnants.  
The distance of the massive dense clump under study is not well established as it lies on the tangential direction 
of the spiral arm or bridge structure to which the Sun belongs in our Galaxy \citep{sha65,ker65}.  
Many studies reporting on observations of IRAS 20126+4104 
adopt a distance of 1.7 kpc, based on the value quoted by \citet{dam85}.
However, \citet{dam85} do not explain how such a distance was derived.
The kinematical estimate using a galactic rotation model may not be accurate for this source because of the location of 
the source relative to the Galactic Center.  
An alternative method is that of looking for a possible interaction of the IRAS 20126+4104 clump
with nearby objects whose distances are known.  
 From the north-east to the north of the source, there is a chain of \HII\ regions,
whose northern edge coincides with IC1318a, the biggest HII region in the cluster.
IRAS 20126+4104 is located close to the south-western edge of the chain.  
The angular distance from the closest HII region 
is about 7\arcmin.  
In addition, there is a number of HII regions to the east of the source
and one supernova remnant (SNR), G 78.2+2.1 \citep{hig77,lan80} 
with diameter of $\sim$62\arcsec~ to the south-east of it.
The angular distance between the dense clump and the supernova remnant is about 1.5\deg, along a direction with $PA$ $\sim$ 120\deg.
To the south-east of the object, there is an O star, HD193322 (O9V).  
The angular separation between HD193322 and IRAS 20126+4104 is about 51\arcmin.
The distance of the O star is estimated to be $\sim$1 kpc from photometry \citep{cru74}. 
To the south-west, there is no strong HII region. 
Ionized gas appears to surround the massive dense clump 
except on the south-west side, as shown in Figure 1 ($\S$ \ref{sharc}), which is evidence of interaction
between our clump and all these objects.
The ionized gas is likely due to irradiation from these HII regions and from the SNR. 
As the distance of these sources is reported to be 1.5 $\pm$ 0.5 kpc
(Iksanov 1960, Dickel et al. 1969, Higgs et al. 1977, Landecker et al. 1980, Piepenbrink \& Wendker 1988, and references therein),
in this paper we assume a distance of 1.5 kpc for IRAS 20126+4104.

\clearpage
\begin{deluxetable}{lrrrrrrcr}
\tabletypesize{\scriptsize}
\rotate 
\tablecaption{Observations \label{tbl-1}}
\tablewidth{0pt}
\tablehead{
\colhead{Emission} & \colhead{Frequency} & \colhead{Bandwidth} & \colhead{Telescope/}   & \colhead{$\theta_{\rm HPBW}$\tablenotemark{a} }  & \colhead{FOV\tablenotemark{b} }  & \colhead{Velocity resolution}  & 
\colhead{Reference}  \\
\colhead{}     & \colhead{(GHz)} & \colhead{(MHz)} & \colhead{Instrument}   & \colhead{(\arcsec)}  & \colhead{(arcmin$^{2}$)}  & \colhead{(kms$^{-1}$)} & 
\colhead{} 
}
\startdata
350$\mu$m continuum & 849.4 &102900  &CSO/SHARC II  & 8.7 & 9 $\times$ 9 & ... &  This work \\
450$\mu$m continuum & 658.3 & 67900  &CSO/SHARC II  & 9.8 & 9 $\times$ 9 & ... &  This work \\
CO $J=6-5$ & 691.473076 & 50 & CSO/690GHz Rx  &10  & 2.7 $\times$ 2.7 & 0.6 & This work \\
CO $J=7-6$ & 806.651776 & 250 & HHT\tablenotemark{c}/800GHz Rx  &10  &...             & 0.9 & \citet{kaw99}\\
 \enddata


\tablenotetext{a}{Half power beam width.}
\tablenotetext{b}{Field of view.}
\tablenotetext{c}{The Heinrich Hertz Telescope.}
\tablenotetext{d}{The largest angular scale detected with the interferometric observations = 16\arcsec.} 


\end{deluxetable}

\clearpage

\begin{deluxetable}{crrrrcr}
\tabletypesize{\scriptsize}
\tablecaption{Peak Flux Densities and Total Flux \label{tbl-2}}
\tablewidth{0pt}
\tablehead{
\colhead{Wavelengths} & \colhead{Telescope/}   & \colhead{$\theta_{\rm HPBW}$\tablenotemark{a} }  & \colhead{Flux density\tablenotemark{b}} 
& \colhead{Total flux\tablenotemark{b}}   &  \colhead{Effective radius}   & \colhead{Reference}  \\
\colhead{($\mu$m)}    & \colhead{Instrument}   & \colhead{(\arcsec)} & \colhead{(Jy Beam$^{-1}$)}  & \colhead{(Jy)}   & \colhead{(\arcsec)}  & \colhead{} \\
}
\startdata
350  & CSO/SHARC II & 8.7  & 59.1 (4.1)  & 477 (13) & 65.5 &  This work \\
350  & CSO/SHARC    & 11  & 34              &  130          &  25 &   \citet{hun00} \\
450  & CSO/SHARC II  & 9.8 & 27.4 (1.2) &  137 (14)  & 50.4 & This work \\
450  & JCMT/SCUBA  & 8    & 29 (0.9)     &    91.1(3.9)   & ...  &  \citet{wil04} \\
450  & JCMT/SCUBA  & 12  & 32.0 (6.0)  & 162 (24) & ... & \citet{ces99} \\
850  & JCMT/SCUBA  & 14.4 & 5.6 (0.04)  &  21.9 (0.5) & 15  &\citet{wil04} \\
850  & JCMT/SCUBA  & 21   & 5.3 (0.6)  & 19 (3)       & ... &\citet{ces99} \\
1200& Pico Veleta/MAMBO  & 11   & 1.1 (0.3)  & 5.8 & 20 & \citet{beu02} \\
1350& JCMT/SCUBA  & 22   & 1.74 (0.20)  & ... & ... &\citet{ces99} \\
2000& JCMT/SCUBA  & 33  & 0.56 (0.05)  & ... & ... & \citet{ces99} \\
 \enddata


\tablenotetext{a}{Half power beam width.}
\tablenotetext{b}{Error is given in the parenthesis when the literatures provide. }


\end{deluxetable}


\clearpage
\begin{deluxetable}{crrrrrrrrr}
\tabletypesize{\scriptsize}
\rotate 
\tablecaption{IRAS 20126+4104 CO Outflow Parameters  \label{tbl-3}}
\tablewidth{0pt}
\tablehead{
\colhead{Transition} & 
\multicolumn{2}{c}{$l$\tablenotemark{a}} &
\multicolumn{2}{c}{$M$\tablenotemark{b}} &
\colhead{$PA$}  & 
\multicolumn{2}{c}{$V$\tablenotemark{c}} &
\colhead{$\tau_{\rm dyn}$} &
\colhead{Reference}  \\
\cline{2-3}
\cline{4-5}
\cline{7-8}
\colhead{} & 
\colhead{Red} & 
\colhead{Blue} & 
\colhead{Red}& 
\colhead{Blue}  & 
\colhead{}  & 
\colhead{Red}  & 
\colhead{Blue} &
\colhead{}  & 
\colhead{} \\
\colhead{} & 
\colhead{(pc)} & 
\colhead{(pc)} & 
\colhead{(\Msun)}& 
\colhead{(\Msun)}  & 
\colhead{(\deg)}  & 
\colhead{(\kms)}  & 
\colhead{(\kms)} &
\colhead{(yr)} &
\colhead{} 
}
\startdata
$J=6-5$ & 0.23& 0.16 & 13 & 28 & 50 & 35 & 30  & $\gtrsim$ 1.7 $\times 10^{4}$  & This work \\
$J=1-0$ & 0.64 & 0.35 & 20 & 33 & -10 & 25 & $>$ 28  & 6.4$\times 10^{4}$ & \citet{she00}\\ 
 \enddata

\tablenotetext{a}{Outflow lobe radius.}
\tablenotetext{b}{Mass of the outflow lobes.}
\tablenotetext{c}{Outer boundary of the wing components.}
\tablenotetext{d}{Dynamical timescale.}


\end{deluxetable}





\clearpage

\clearpage

\clearpage

\clearpage

\clearpage

\clearpage

\clearpage

\clearpage

\clearpage

\clearpage
\clearpage

\end{document}